\documentclass[prl,aps,showpacs,twocolumn]{revtex4}

\usepackage{times}
\usepackage{graphicx}
\usepackage{float}
\usepackage{latexsym,amsmath,amssymb,bm,euscript}
\usepackage{color}
\usepackage{subfigure}
\usepackage{epstopdf}
\usepackage[colorlinks=true,linkcolor=blue,citecolor=blue]{hyperref}

\begin{document}

\title{Topological Characterization of Non-Abelian Moore-Read State using Density-Matrix Renormailzation Group}
\author{W. Zhu$^1$, S. S. Gong$^1$, F. D. M. Haldane$^2$ and D. N. Sheng$^1$}
\affiliation{$^1$Department of Physics and Astronomy, California State University, Northridge, California 91330, USA}
\affiliation{$^2$Department of Physics, Princeton University, Princeton, NJ 08544, USA}

\begin{abstract}
The non-Abelian topological order has attracted a lot of attention  for its fundamental importance
and  exciting prospect of topological quantum computation. However, explicit demonstration or identification
of the non-Abelian states and the associated statistics in a microscopic model is very  challenging.
Here, based on density-matrix renormalization group calculation,
we provide a complete  characterization of the universal properties of
bosonic Moore-Read state on Haldane honeycomb lattice model at filling number $\nu=1$ for larger systems,
including both the edge spectrum and the bulk anyonic quasiparticle (QP) statistics.
We first demonstrate that there are three degenerating ground states,
for each of which there is a definite anyonic flux threading through the cylinder.
We identify the nontrivial countings for the entanglement spectrum
in accordance with the corresponding conformal field theory.
Through simulating a flux-inserting experiment, it is found that two of the Abelian ground states can be
adiabatically connected, while the ground state in Ising anyon sector evolves back to itself,
which reveals the fusion rules between different QPs in real space.
Furthermore, we calculate the modular matrices $\mathcal{S}$ and $\mathcal{U}$,
which contain all the information for the anyonic QPs such as quantum dimensions, fusion rule and topological spins.
\end{abstract}

\maketitle


\textit{Introduction.---}
The topological order of a quantum state is correlated with the pattern of long-range quantum entanglement \cite{Wen1990,XChen2010},
which is characterized by ground state (GS) degeneracy on compactified space \cite{Wen1989},
gapless edge states \cite{Wen1991,Wen1992,Wen1995},
and fractional quasiparticles (QPs) with anyonic statistics.
According to the braiding statistics of QPs, the topological ordered states
are generally classified as
Abelian \cite{Laughlin} and non-Abelian \cite{Moore,Greiter,Read} states.
An interchange of two Abelian QPs leads to a nontrivial phase acquired by
their wavefunction. On the other hand, an interchange of two non-Abelian QPs
transforms the system from one GS to another and the final state
will depend on the order of the implemented operations.
The non-Abelian QPs and their  braiding statistics
are essential  information  for understanding the topological
order, which can also lead to potential applications in topological
quantum computation \cite{Kitaev2003,Sarma,Nayak}.

\begin{figure*}[t]
 \begin{minipage}{0.98\linewidth}
 \centering
 \includegraphics[width=6.8in]{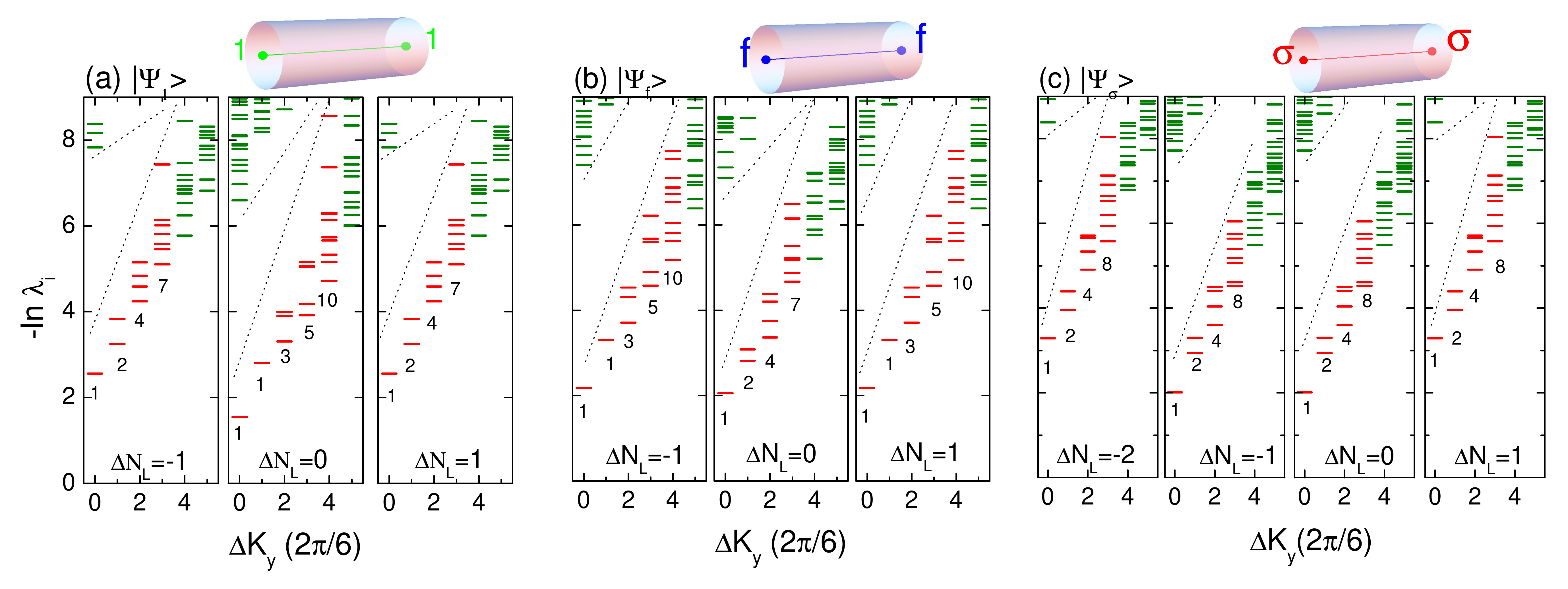}
 \end{minipage}
\caption{(Color online) The ES for three topological GSs: (a) Identity GS $|\Psi_{\openone}\rangle$, (b) Fermion GS $|\Psi_{f}\rangle$ and (c)Ising anyon GS $|\Psi_{\sigma}\rangle$,
where $\lambda_i$ is the eigenvalue of reduced density matrix $\hat{\rho}_L$ of left half of an infinite cylinder.
The ES are labeled by the relative boson number $\Delta N_L=N_L-N_L^0$ of left half cylinder in each tower
($N_L^0$ is the boson number of the state of $\hat{\rho}_L$  with the largest eigenvalue).
In each tower, the horizontal axis shows the relative momentum $\Delta K_y=K_y-K_y^0$
in the transverse direction of the corresponding eigenvectors of $\hat{\rho}_L$
($K_y^0$ is momentum of the state with the largest eigenvalue for  $\hat{\rho}_L$ in each tower).
The numbers below the red dots label the nearly degenerating pattern for the low-lying ES with different  $\Delta K_y$.
The black dashed line shows the entanglement gap in each momentum sector.
Here the calculation is performed on $L_y=6$ cylinder using infinite DMRG with keeping $3200$ states.
}\label{ES}
\end{figure*}

Identifying and characterizing the emergent topological order
in strongly correlated systems has been regarded as a very challenging task.
Recently, quantum entanglement has been extensively used to describe the emergent
topological order in strongly interacting systems \cite{Kitaev,Levin,Haldane2008,YZhang2012,SDong,Cincio_PRL,Zaletel_PRL,Tu2013},
which has also offered a new route for characterizing the topological order by obtaining the modular matrices of the  systems.
In particular, two theoretical progresses are prominent for uniquely identifying  a topological order.
First, Li and Haldane have established that
 the entanglement spectrum (ES)
of groundstate of fractional quantum Hall state
contains information about their edge modes \cite{Haldane2008},
if no edge reconstruction occurs in the system \cite{Chandran2014}.
Since the gapless edge state is universal for topological ordered systems
governed by  the conformal field theory (CFT) \cite{Wen1995},
the ES provides a fingerprint of the topological order.
Secondly,  Zhang \textit{et al} demonstrated that the braiding statistics of anyonic QPs
can be extracted from the minimal entangled states by constructing the modular matrices using the projected variational wavefunctions.
Theoretically, this is based on the fact that the minimal entangled state is an  eigenstate of the Wilson loop operator
with a definite type of QP \cite{SDong}, which can be used as the basis states for modular transformation.
This approach  has been applied to characterize different Abelian topological ordered states including
 the Laughlin states on topological bands\cite{Cincio_PRL,WZhu2013},
chiral spin liquid \cite{SSGong_SR,YCHe2014,Bauer} and $Z_2$ spin liquid \cite{WZhu_JSM} on extended spin$-1/2$ kagome lattice
models based on exact diagonalization (ED) or large scale density matrix renormalization group (DMRG) simulations\cite{White_PRL,McCulloch}.
Since non-Abelian QPs are much more interesting
and their properties are richer and significantly different from the Abelian ones,  it is
highly desired to extract the non-Abelian statistics using such kind of entanglement measurement.
However, due to the limited computational capability,   
only partial information such as mutual statistics 
has been successfully obtained in previous studies \cite{YZhang2013,WZhu2014},
which is not sufficient to uniquely classify a non-Abelian topological ordered state.
Taking the non-Abelian Moore-Read state as an example,
there are $8$ different related chiral Ising CFTs that share the same mutual statistics \cite{Kitaev2006}.
To distinguish them, one needs the self statistics of QPs or chiral central charge \cite{Kitaev2006}. 
Therefore, both mutual and self statistics are necessary to determine a non-Abelian topological order \cite{Tu2013,YZhang2014,note2},
which requires  an unbiased numerical method to obtain topological degenerating quantum states
for larger systems  and  overcome the limitation of previous methods.

The topological ordered state has fractionalized QPs.  One intrinsic property of  the anyonic QPs is
the fusion rule of the QPs that a combination of two anyonic QPs yields one or more than one type of different QPs,
which is the fundamental concept for future qubit-based topological quantum computation.
Although the fusion rules can be alternatively obtained from modular $\mathcal{S}$ matrix through the Verlinde formula \cite{Verlinde},
it is also highly desired to directly demonstrate the fusion process between two given anyonic QPs in \textit{real space}.
The  simulation of  the QP fusion rule is regarded as a very difficult  task  and has not been directly demonstrated
for microscopic non-Abelian systems.   
Recently, we\cite{SSGong_SR} illustrate a method of combining Laughlin gedanken experiment \cite{Laughlin1981} and
ES measurement \cite{Haldane2008} to simulate the QP fusion rule.
To generalize this method from Abelian system \cite{SSGong_SR} to
non-Abelian system is another goal of the current work.

The aim of this paper is to
provide compelling numerical evidences of the non-Abelian nature of bosonic Moore-Read state
in a microscopic lattice model for large systems.
Based on the DMRG calculations, we are able to access  a complete set of topological GSs with different anyonic flux threading through the cylinder,
which can be identified by the characteristic edge spectrum governed by $SU(2)_2$ CFT.
Then we apply the newly developed adiabatic DMRG to this system \cite{SSGong_SR}.
By adiabatically threading a $U(1)$ charge flux, it is found that the two Abelian GSs can be
adiabatically connected through pumping a QP with unit charge from one edge to the other,
while the non-Abelian GS only evolves back to itself.
Importantly, this pumping and transferring QP process is equivalent to the simulation of fusion rules between different QPs.
To our best knowledge, this is the first time to demonstrate such kind of fusion rules of non-Abelian system in real space.
Moreover, using the GSs in all topological sectors, we also calculate the modular $\mathcal{S}$ and $\mathcal{U}$ matrices.
which contain the mutual and self statistics of all three kinds of  QPs.
On one hand, the fusion rules from the modular $\mathcal{S}$ matrix self-consistently validates the flux insertion simulation.
On the other hand, the further information (i.e. topological spin, central charge) from modular $\mathcal{U}$ matrix
helps us determine Haldane honeycomb model realizes $n=1$ chiral Ising theory \cite{Kitaev2006}.

\textit{Model and method.---}
We study the Haldane model on the honeycomb lattice \cite{Haldane1988} filled with interacting bosons: 
\begin{eqnarray}
&&H=t\sum_{\langle\mathbf{r}\mathbf{r}^{\prime}\rangle}
\left[b^{\dagger}_{\mathbf{r}^{\prime}}b_{\mathbf{r}}+\textit{h.c.}\right]
+t^{\prime}\sum_{\langle\langle\mathbf{r}\mathbf{r}^{\prime}\rangle\rangle}
\left[b^{\dagger}_{\mathbf{r}^{ \prime}}b_{\mathbf{r}}e^{i\phi_{\mathbf{r}^{\prime}\mathbf{r}}}+\textit{h.c.}\right]\nonumber\\
&&+t^{\prime\prime}\sum_{\langle\langle\langle\mathbf{r}\mathbf{r}^{
\prime}\rangle\rangle\rangle}
\left[b^{\dagger}_{\mathbf{r}^{\prime}}b_{\mathbf{r}}+\textit{h.c.}\right]
+\sum_{n}\frac{U_n}{n!}\sum_{\mathbf{r}}(b^{\dagger}_{\mathbf{r}})^n (b_{\mathbf{r}})^n, \label{hamilton}
\end{eqnarray}
where $b^{\dagger}_{\mathbf{r}} (b_{\mathbf{r}})$
creates (annihilates) a boson at site $\mathbf{r}=(x,y)$.
Here, we adopt the parameters of the nearest neighbor (NN) hopping $t=-1$,
the second NN $t^{\prime}=-0.60$ and $\phi=0.4\pi$, the third NN
$t^{\prime\prime}=0.58$ \cite{YFWang2012}.
We also set the on-site \textit{N}-body repulsive interaction as
$U_{2}=0$ and $U_{n>2}=\infty$, which is equivalent to the
``three-body hard-core boson'' condition:
$\left(b^{\dagger}_{\mathbf{r}}\right)^3=0$ and
$\left(b_{\mathbf{r}}\right)^3=0$ \cite{Cirac2007}.
This model can also be considered as a spin-1 model through the standard mapping \cite{Cirac2010}.
In this paper we focus on the filling factor $\nu=1$ for studying the interesting Moore-Read state\cite{YFWang2012}.

In this work, we study the Hamiltonian Eq. (\ref{hamilton}) on cylinder geometry
using the infinite DMRG combined with finite DMRG method \cite{White_PRL,McCulloch,Cincio_PRL,Zaletel_PRL}.
We consider the cylinders with finite width $L_y=4,6$ (measured by the number of unit cells).
We have kept up to $3200$ states in the DMRG simulation.
The different topological GSs are obtained by the random boundary condition \cite{Cincio_PRL},
targeting the excited state in the initial process, and the inserting flux method\cite{YCHe2013}.
The DMRG is especially efficient to deal with the topological ordered and gapped system, which
allows us to obtain the GS with well-defined anyonic flux \cite{Cincio_PRL,HCJiang_Nat,Poilblanc2013,YCHe2013}.
Compared to the ED calculations \cite{YFWang2012,WZhu2014},
the DMRG algorithm offers great advantages because it can access larger system sizes accurately.
More importantly, by implementing
the state-of-art techniques for  detecting topological order  in DMRG simulations,
we can identify and characterize the topological nature of a potential topological ordered state
in an interacting  system, both at the edge and in the bulk.

\textit{Chiral edge spectrum.---}
Through initializing the boundary condition, targeting excited state for non-abelian sector,  and optimizing the bulk of the cylinder\cite{Cincio_PRL},
we obtain three  nearly degenerated GSs (the bulk energy difference per site is less than $0.0004$).
We anticipate that the three GSs host
distinct and well-defined topological sectors  with different flux $\textit{a}$  through the cylinder.
From the chracateristic ES discussed below (see Fig. \ref{ES}),
these sectors can be identified as the identity $\textit{a}=\openone$, fermion $\textit{a}=f$ and Ising anyon $\textit{a}=\sigma$ sectors.
When the cylinder is being cut into two halves, a $\textit{a}-$type QP appears near the edge of the cut,
which leads to different gapless edge excitation that can be distinguished by the ES.

Fig. \ref{ES} shows the ES for each of the three GSs $|\Psi_{a}\rangle$ ($\textit{a}=I,f,\sigma$) obtained on a cylinder.
The ES is  grouped by the relative boson number $\Delta N_L$ of the half system
and their relative momentum quantum number $\Delta K_y$ (both $\Delta N_L$ and $\Delta K_y$ are related
to the quantum numbers of the lowest level in ES without flux for each topological  sector) along the transverse direction (referred to as y-direction).
In Fig. \ref{ES}(a), the leading ES of $|\Psi_{\openone}\rangle$ displays the sequence of degeneracy pattern $\{1,1,3,5,10,...\}$ in even $\Delta N_L$ sector and $\{1,2,4,7,...\}$ in odd $\Delta N_L$ sector.
This even-odd effect can be understood from the root configuration $``..02020202..''$ of $|\Psi_{\openone}\rangle$ depending on the microscopic environment near a cut \cite{Bernevig2008,Bergholtz2005,Bernevig2012,Ardonne2005,Papic,ZLiu2012}.
Importantly, the edge mode countings agree with the prediction of the identity primary field and its descendants in $SU(2)_2$ Wess-Zumino-Witten CFT.
Similarly, as shown in Fig. \ref{ES}(b), the low-lying ES of $|\Psi_{f}\rangle$ shows degeneracy pattern $\{1,2,4,7,...\}$ in even $\Delta N_L$ sector and $\{1,1,3,5,10,...\}$ in odd $\Delta N_L$ sector, as expected from the fermion primary field and its descendants.
Physically, the $|\Psi_{f}\rangle$ is equivalent to $|\Psi_{\openone}\rangle$ with creating a pair of charge $\textit{e}$ ($\textit{e}$ is the unit charge) QPs at two ends of the cylinder (see below).
Therefore, the even-odd effect in the ES of $|\Psi_{f}\rangle$ is shifted by $\Delta N_L=1$, compared to that of $|\Psi_{\openone}\rangle$.
Next we turn to results  of the ES of $|\Psi_{\sigma}\rangle$, as shown in Fig. \ref{ES}(c),
which shows two significant differences compared to the $|\Psi_{\openone}\rangle$ and $|\Psi_{f}\rangle$.
First, the ES is symmetric about $\Delta N_L=-1/2$ rather than $\Delta N_L=0$.
This feature results from the Ising anyon QP $\sigma$ created at each edge of the cylinder,
carrying the  fractional charge $\textit{e}/2$.
Second, the ES shows the same degeneracy pattern $\{1,2,4,8,...\}$ in all $\Delta N_L$ sectors.
It can be understood from the root configuration $``..11111111..''$ of $|\Psi_{\sigma}\rangle$ \cite{ZLiu2012}.
These observations are consistent with the analytical prediction of Ising anyon primary field according to $SU(2)_2$ CFT \cite{Ardonne2005}.

\begin{figure}[t]
 \begin{minipage}{0.45\linewidth}
 \centering
 \includegraphics[width=1.6in]{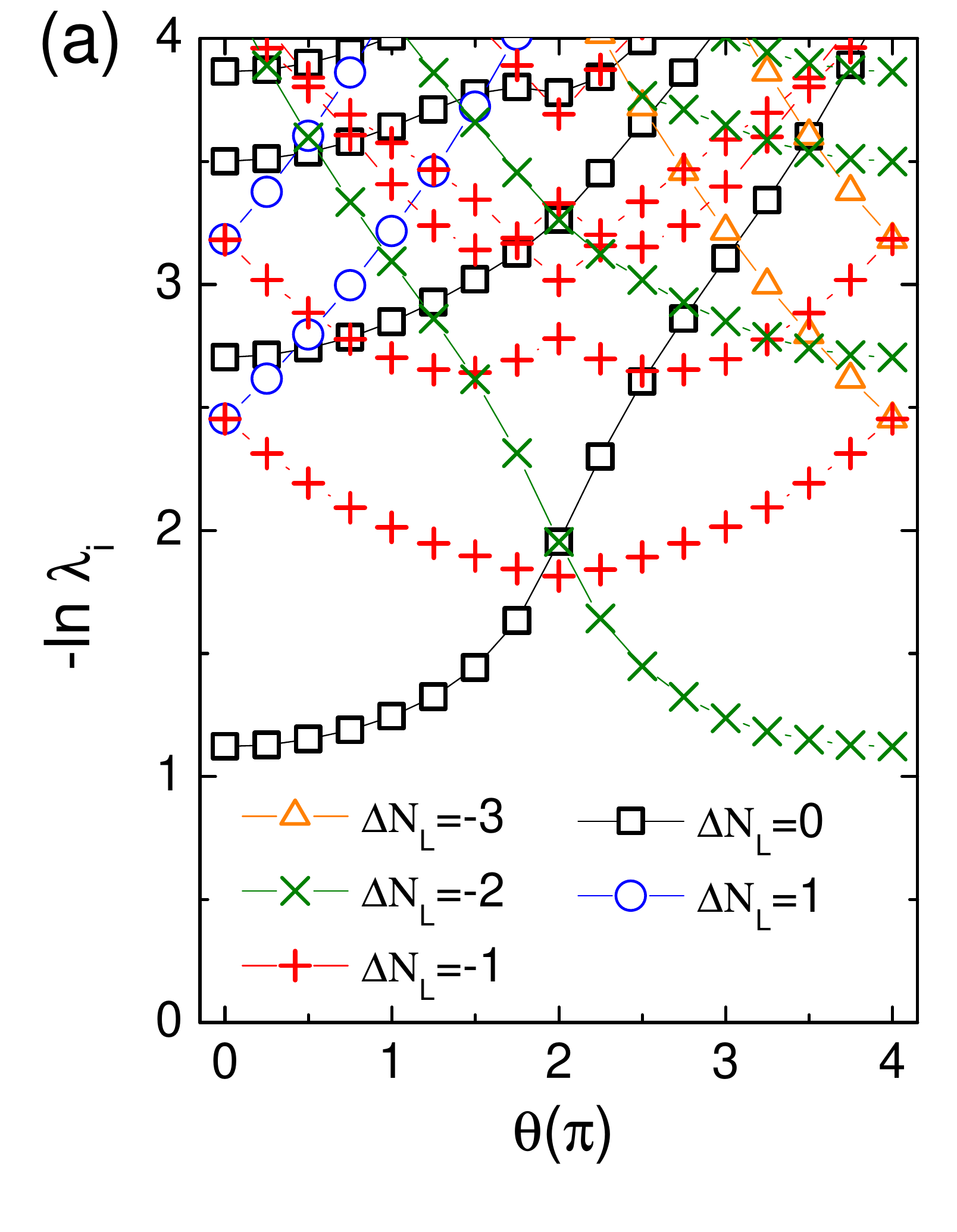}
 \end{minipage}
 \begin{minipage}{0.45\linewidth}
 \centering
 \includegraphics[width=1.6in]{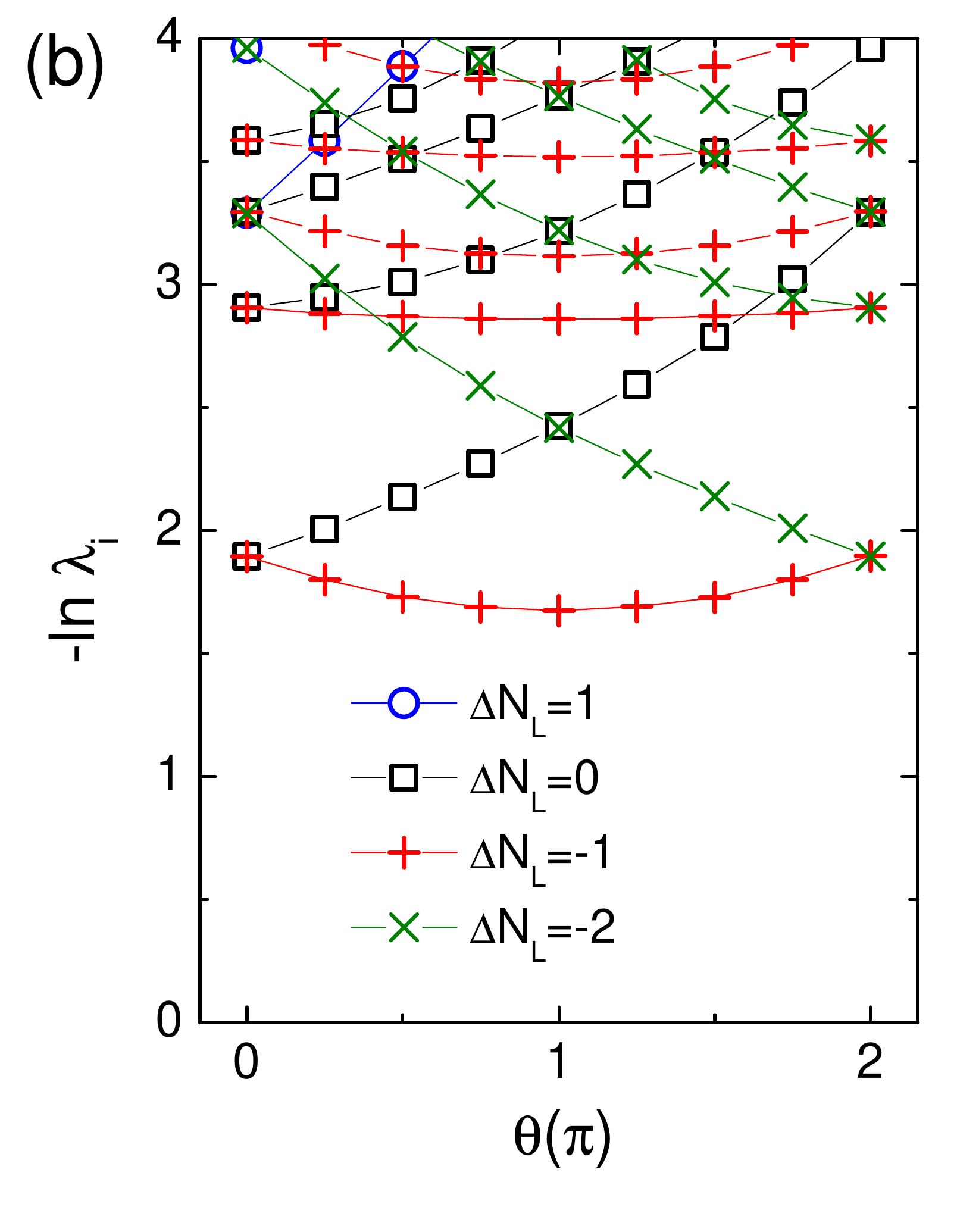}
 \end{minipage}
\caption{(Color online) The ES flow with inserting flux $\theta$ in the hole of the cylinder:
(a) Starting from the GS $|\Psi_{\openone}\rangle$ at $\theta=0$ and adiabatically threading a $\theta=4\pi$ flux.
(b) Starting from the GS $|\Psi_{\sigma}\rangle$ and threading a $2\pi$ flux.
Here the calculation is performed on $L_y=6$ cylinder using infinite DMRG with keeping $1200$ states.
}\label{flux}
\end{figure}

\textit{Flux insertion.---}
We further perform the numerical flux insertion simulations on cylinder systems based on the newly developed adiabatic DMRG \cite{SSGong_SR,YCHe2013,Zaletel_flux,WZhu_JSM,Grushin2014,WZhu_CSL}.
Due to the quantized Hall response \cite{Laughlin1981},
it is expected that a quantized charge will be pumped from left edge to the right edge by inserting a $U(1)$ charge flux. 
The dynamical pumping process reveals the nature of the pumped QP and 
the fusion rules between different QPs.

\begin{figure}[t]
 \begin{minipage}{0.99\linewidth}
 \centering
 \includegraphics[width=3.5in]{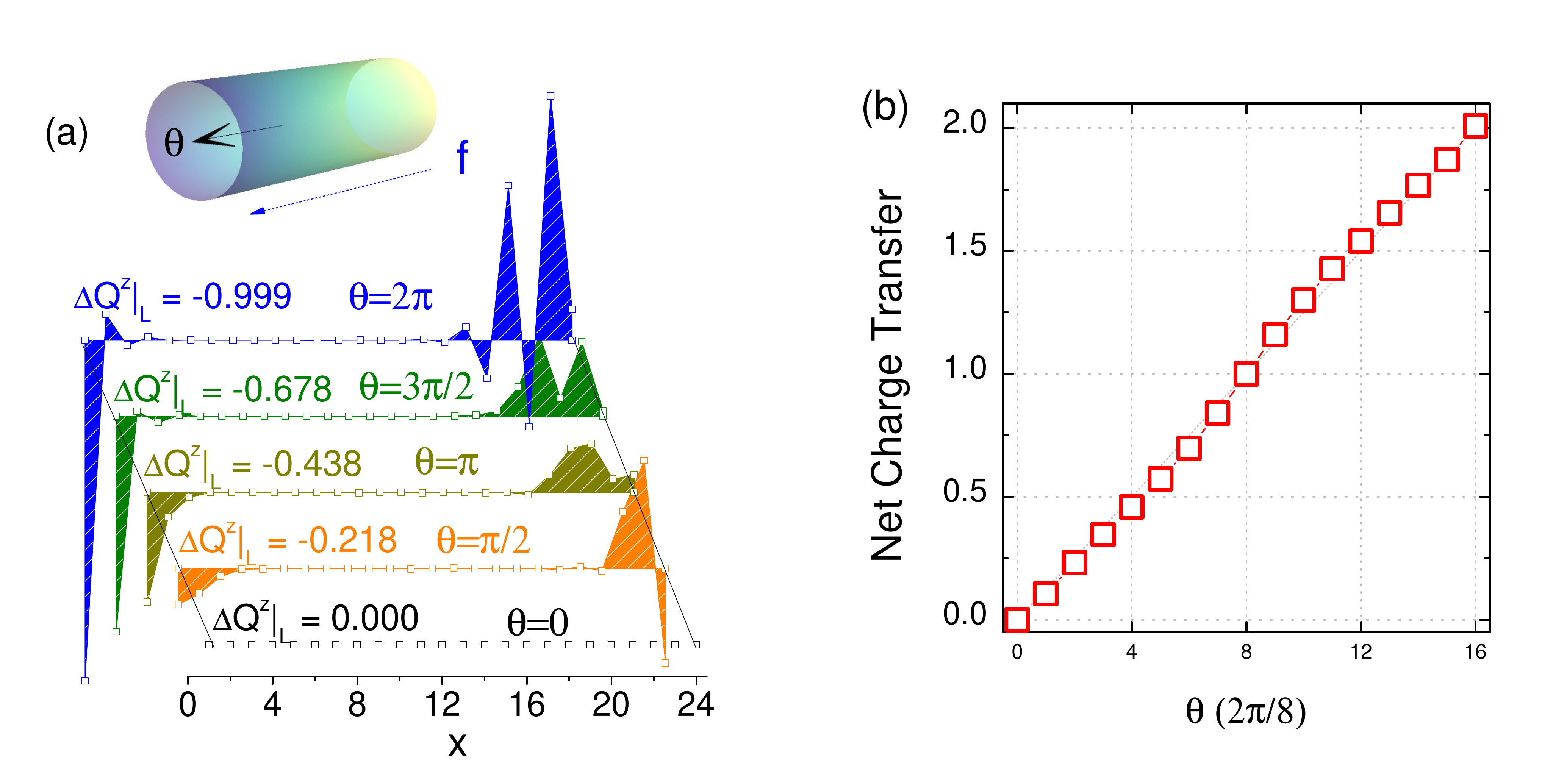}
 \end{minipage}
\caption{(Color online) Real-space configuration of the accumulated charge $\langle \Delta Q_x\rangle = \sum_{y}\langle \Delta Q_{x,y}\rangle$
(the summation is over all the $2L_y$ sites in each column $x$) with increasing flux $\theta$. $\Delta Q_{L(R)}$ is defined by the total charge localized on left (right) end of cylinder. Here the calculation is performed on $L_y=4$ cylinder with length $L_x=24$.}\label{Net_charge}
\end{figure}

As shown in Fig. \ref{flux} (a), by threading a $2\pi$ flux, $|\Psi_{\openone}\rangle$ ($|\Psi_{f}\rangle$) adiabatically evolves into $|\Psi_{f}\rangle$ ($|\Psi_{\openone}\rangle$).
Further increasing flux up to $4\pi$ will drive the system  back to the $|\Psi_{\openone}\rangle$ ($|\Psi_{f}\rangle$).
Interestingly, comparing the ESs at $\theta=0$ and $2\pi$, the adiabatic flux insertion shifts the lowest level of ES from $\Delta N_L=0$ to $\Delta N_L=-1$,
signaling a unit charged $f$ QP transferred from left edge to right edge.
Alternatively, we can visualize the charge transferring mechanism from the charge accumulation in real-space.
As shown in Fig. \ref{Net_charge}(a), with adiabatically threading a flux quantum, a net charge accumulation develops at left edge from $\Delta Q_L=0.0$ at $\theta=0$ to $\Delta Q_L=-0.999$ at $\theta=2\pi$. At the right edge, the charge accumulation $\Delta Q_R=-\Delta Q_L$ always holds because of the particle number conservation. In fact, by inserting a single flux quantum, a net charge transfer from left edge to right edge is $\Delta Q=\Delta Q_L=-\Delta Q_R\approx 1.0$ (in the units of charge
 quantum $\textit{e}$).
If inserting two flux quanta, a net charge transfer $\Delta Q=2.0$ is expected (Fig. \ref{Net_charge}(b)),
and consequently  the ES evolves back with the quantum number of ES $\Delta N_L$ shifted by $2$  as shown  in Fig. \ref{flux} (a).
In this process, two $f$ QPs are pumped from one edge to another, and they combine,
which drives the bulk GS $|\Psi_{\openone}\rangle$ ($|\Psi_{f}\rangle$) back to itself.
Thus we find the $|\Psi_{\openone}\rangle$ ($|\Psi_{f}\rangle$) hosts even (odd) number of edge $f$ QPs,
which are two independent Abelian sectors of the system.
More importantly, pumping and transferring QP from one edge to the other edge actually simulates the QP fusion process:
Threading a $2\pi$ and $4\pi$ flux respectively relates to $\openone\times f=f$ and $f\times f=\openone$.
There are similar to the $\nu=1/2$ Laughlin state but the $f$ QP here carries
unit charge $\textit{e}$ and they also satisfy different self statistics (see below).

Interestingly, as shown in Fig. \ref{flux}(b),
the Ising anyon GS $|\Psi_{\sigma}\rangle$ will evolve into itself by threading a flux quantum,
although a net charge $\textit{e}$  QP transfer occurs.
It directly results from the fusion rule of Ising anyon $\sigma$ QP: To combine one $\sigma$ QP (charge-$\textit{e}/2$) and one $f$ QP (charge-$\textit{e}$) is equivalent to one $\sigma$ QP : $\sigma\times f = \sigma$.
Moreover, the Ising anyon $\sigma$ QP does not respond to the $U(1)$ charge flux,
which is significantly different from the charged fermion $f$ QP. 

\textit{Anyonic statistics of QPs.---}
The braiding  statistics of anyonic QPs are encoded in the
modular $\mathcal{S}$ and $\mathcal{U}$ matrices \cite{Wen1990,SDong,Verlinde,ZHWang,Fendley,Bonderson,Ardonne,Bais}.
In the topological quantum field theory,
the modular matrices describe the action of  modular transformation
on the eigenstates of the Wilson loop operators representing  different types of QPs.
Because the QP eigenstates always select the minimal entropy \cite{SDong},
one can  use the minimal entangled states as the canonical basis for defining $\mathcal{S}$ and $\mathcal{U}$\cite{YZhang2012,WZhu2013,WZhu2014}.
Remarkably,  the minimal entangled states are these states we obtain in DMRG as described above.

Following the procedure outlined in Ref. \cite{Cincio_PRL,Sandvik}, we obtain the modular matrix
on $L_y=4$ cylinder: 
\begin{eqnarray*}
\mathcal{S} \approx \mathcal{S}^{CS}+
\left[\begin{array}{ccc}
0.068 & -0.035 & 0.015 \\
-0.019 & -0.049+0.150i & 0.039+0.146i \\
0.036 & 0.061+0.121i &  0.050+0.077i \\
\end{array}\right]
\end{eqnarray*}
and
\begin{eqnarray*}
\mathcal{U} \approx \mathcal{U}^{CS}\times
\left[\begin{array}{ccc}
e^{i0.029\pi} & 0 & 0 \\
0 & e^{-i0.083\pi} & 0 \\
0 & 0 & e^{-i0.041\pi} \\
\end{array}\right].
\end{eqnarray*}
Indeed, the numerical obtained modular matrices are quite close to the analytical prediction
from $SU(2)_2$ Chern-Simons theory  \cite{SDong,ZHWang,Fendley}:
$\mathcal{S}^{CS} = \frac{1}{2}\left[\begin{array}{ccc}
1 & 1 & \sqrt{2} \\
1 & 1 & -\sqrt{2} \\
\sqrt{2} & -\sqrt{2} & 0 \\
\end{array}\right]$ and
$\mathcal{U}^{CS} = e^{-i\frac{2\pi}{24}\frac{3}{2}}
\left[\begin{array}{ccc}
1 & 0 & 0\\
0 & -1 & 0\\
0 & 0 & e^{i3\pi/8}
\end{array}\right]$.

In general, from the modular matrices, we have the full statistics information of emerging   QPs:
i) The  QPs Identity ($\openone$), fermion ($f$) and Ising anyon ($\sigma$)  have the quantum dimensions \cite{Kitaev,Levin} $d_{\openone}=1$, $d_f=1$,  and $d_{\sigma}=\sqrt{2}$ respectively and the total quantum dimension is $\mathcal{D}=2$.
ii) The fusion rule of QPs (that specifies how
the QPs combine and fuse) \cite{ZHWang,Fendley,Bonderson,Ardonne}: $\openone\times x=x$ ($x=\openone,f,\sigma$), $f\times f=\openone$, $\sigma\times \sigma=\openone+f$ and $\sigma\times f=f\times\sigma=\sigma$.
iii) The topological spins (from the phase factor for the QP obtained during a self-rotation of $2\pi$): $h_{\openone}=0$, $h_{f}=1/2$ and $h_{\sigma}=3/16$,
respectively.  
iv) The chiral central charge  $c=3/2$.
In particular, the non-trivial quantum dimension $d_{\sigma}=\sqrt{2}$
signals the non-Abelian fusion rule of $\sigma$ QPs:  $\sigma\times \sigma=\openone+f$,
 that two $\sigma$ QPs may either fuse into an $\openone$ or a $f$ QP.
Therefore each pair of Ising anyon QPs can act as a qubit for quantum computation \cite{Nayak}.
Moreover, non-Abelian nature of $\sigma$ QP is also encoded in the topological spin $h_{\sigma}=3/16$, which distinguishes from
boson-like $\openone$ with $h_{\openone}=0$ and fermion-like $f$ with $h_{f}=1/2$.
The topological spin $h_{\sigma}=3/16$ implies
that our model realizes the pure $SU(2)_2$ Chern-Simons theory, rather than the
the $U(4)_1/SU(2)_2$ gauge theory (which has $h_{\sigma}=5/16$) \cite{YZhang2014} or
the non-Abelian state in Kitaev honeycomb model (with $h_{\sigma}=\pm1/16$) \cite{Kitaev2006,note2}.
In addition, the chiral central charge $c=3/2$ further supports
our Haldane honeycomb lattice model realizes the $SU(2)_2$ Chern-Simons theory \cite{Kitaev2006}.
Since the potential non-Abelian phase in Haldane honeycomb lattice model
may be  realized in future cold atom experiments \cite{Jotzu2014},
clarifying which  topological order is realized  in this system provides valuable information
for  the future study.

\textit{Summary and Discussion.---}
We have numerically studied the universal properties of
the non-Abelian Moore-Read state by revealing  both the characteristic ES and the bulk topological nature,
without using the empirical knowledge of model wave functions.
The two GSs with Abelian QPs can be distinguished by transporting a fermionic $f$
QP on a cylinder and the corresponding unit charge of $f$ QP is determined simultaneously through inserting flux simulation.
Interestingly, the QP pumping and transferring process naturally demonstrates the fusion rules in such non-Abelian system in real space.
In addition, extracting the modular matrices from the GSs of DMRG, we justify the completeness of the GSs,
and determine which CFT is realized for the  non-Abelian state of the system.
Interesting systems for future studies include the possible non-Abelian state in the spin-1 system \cite{SSGong_SR, Yao2007,Ronny2009}
 and in the double-layer system
with two Abelian Laughlin states coupled together \cite{preparation}.


\textit{Acknowledgement.---}WZ thanks Y. C. He and L. Cincio for their discussions
about developing infinite DMRG algorithm.
WZ also thanks Y. Zhang for insightful discussion.
This work is supported by the U.S. Department of Energy,
Office of Basic Energy Sciences under grant No. DE-FG02-06ER46305 (WZ, DNS).
This work is also supported by the National Science Foundation through grants
the DMR-1408560 and  DMR-1205734 (SSG), and Princeton MRSEC DMR-1420541 (FDMH).
FDMH also acknowledges support from the W. M. Keck Foundation.
DNS also acknowledges travel support from  DMR-1420541 for visiting Princeton.



\begin{thebibliography}{99}

\bibitem{Wen1990}X. G. Wen, Int. J. Mod. Phys. B {\bf 4}, 239 (1990).
\bibitem{XChen2010} X. Chen, Z. C. Gu, and X. G. Wen,
Phys. Rev. B {\bf 82}, 155138 (2010).

\bibitem{Wen1989}X. G. Wen, Phys. Rev. B {\bf 40}, 7387 (1989).

\bibitem{Wen1991}X. G. Wen, Phys. Rev. B {\bf 43}, 11025 (1991).
\bibitem{Wen1992}X. G. Wen, Int. J. Mod. Phys. B {\bf 6}, 1711 (1992).
\bibitem{Wen1995}X. G. Wen, Advances in Physics {\bf 44}, 405 (1995).




\bibitem{Laughlin}R. B. Laughlin, Phys. Rev. Lett. {\bf 50}, 1395 (1983).

\bibitem{Moore}G. Moore and N. Read, Nucl. Phys. B {\bf 360}, 362 (1991).

\bibitem{Greiter}M. Greiter, X. G. Wen and F. Wilczek, Phys. Rev. Lett. {\bf 66}, 3205 (1991).

\bibitem{Read}N. Read and E. Rezayi, Phys. Rev. B {\bf 59}, 8084 (1999).


\bibitem{Kitaev2003}A. Y. Kitaev, Ann. Phys. {\bf 303}, 2 (2003).

\bibitem{Sarma}S. Das Sarma, M. Freedman and C. Nayak, Phys. Rev. Lett. {\bf 94}, 166802 (2005).

\bibitem{Nayak}C. Nayak, S. H. Simon, A. Stern, M. Freedman and S. D. Sarma, Rev. Mod. Phys. {\bf 80}, 1083 (2008).




\bibitem{Kitaev}A. Kitaev and J. Preskill, Phys. Rev. Lett. {\bf 96}, 110404 (2006).
\bibitem{Levin}M. Levin and X.-G. Wen, Phys. Rev. Lett. {\bf 96}, 110405 (2006).

\bibitem{Haldane2008} H. Li and F. D. M. Haldane, Phys. Rev. Lett. {\bf 101}, 010504 (2008).
\bibitem{Chandran2014}A. Chandran, V. Khemani, S. L. Sondhi, Phys. Rev. Lett. {\bf 113}, 060501 (2014).

\bibitem{YZhang2012}Y. Zhang, T. Grover, A. Turner, M. Oshikawa and A. Vishwanath, Phys. Rev. B {\bf 85}, 235151 (2012).

\bibitem{SDong}S. Dong, E. Fradkin, R. G. Leigha and S. Nowling, JHEP 05, 016 (2008).

\bibitem{Cincio_PRL} L. Cincio and G. Vidal, Phys. Rev. Lett. \textbf{110}, 067208 (2013).

\bibitem{Zaletel_PRL}M. P. Zaletel, R. S. K. Mong, and F. Pollmann, Phys. Rev. Lett. {\bf 110}, 236801 (2013).

\bibitem{Tu2013}Hong-Hao Tu, Yi Zhang, and Xiao-Liang Qi, Phys. Rev. B {\bf 88}, 195412 (2013).


\bibitem{WZhu2013}W. Zhu, D. N. Sheng and F. D. M. Haldane, Phys. Rev. B {\bf 88}, 035122 (2013).

\bibitem{SSGong_SR}S. S. Gong, W. Zhu, and D. N. Sheng,
Sci. Rep. \textbf{4}, 6317 (2014).

\bibitem{YCHe2014} Y. C. He, D. N. Sheng, and Y. Chen,
Phys. Rev. Lett. \textbf{112}, 137202 (2014).

\bibitem{Bauer}B. Bauer, L. Cincio, B. P. Keller, M. Dolfi, G. Vidal, S. Trebst, A. W. W. Ludwig,
Nature Commun. {\bf 5}, 5137 (2014).


\bibitem{WZhu_JSM}W. Zhu, S. S. Gong, and D. N. Sheng, J. Stat. Mech. 2014, P08012 (2014).

\bibitem{White_PRL} S. R. White, Phys. Rev. Lett. \textbf{69}, 2863 (1992).
\bibitem{McCulloch} I. P. McCulloch, arXiv:0804.2509.
%
\bibitem{YZhang2013}Y. Zhang and A. Vishwanath, Phys. Rev. B {\bf 87}, 161113(R) (2013).

\bibitem{WZhu2014}W. Zhu, S. S. Gong, F. D. M. Haldane, and D. N. Sheng, Phys. Rev. Lett. {\bf 112}, 096803 (2014).


%
\bibitem{Kitaev2006}A. Y. Kitaev, Ann. Physics {\bf 321}, 2 (2006).

\bibitem{YZhang2014}Y. Zhang and X. L Qi, Phys. Rev. B {\bf 89}, 195144 (2014).

\bibitem{note2}L. Cincio, talk ``Characterizing the topological order by studying the ground states in an infinite cylinder''
at ``Topological Phases of Matter Workshop'', Stonebrook, 06/14/2013.

\bibitem{Verlinde}E. Verlinde, Nucl. Phys. B {\bf 300}, 360176 (1988).


\bibitem{Laughlin1981}R. B. Laughlin, Phys. Rev. B {\bf 23}, 5632(R) (1981)


\bibitem{Haldane1988} F. D. M. Haldane, Phys. Rev. Lett. {\bf 61}, 2015 (1988).
\bibitem{YFWang2012}Y. F. Wang, H. Yao, Z. C. Gu, C. D. Gong, and D. N. Sheng, Phys. Rev. Lett. {\bf 108}, 126805 (2012).


\bibitem{Cirac2007} B. Paredes, T. Keilmann, and J. I. Cirac, Phys. Rev. A {\bf 75}, 053611 (2007).
\bibitem{Cirac2010} L. Mazza, M. Rizzi, M. Lewenstein, and J. I. Cirac, Phys. Rev. A {\bf 82}, 043629 (2010).




\bibitem{HCJiang_Nat}H. C. Jiang, Z. H. Wang and L. Balents, Nat. Phys. {\bf 8}, 902 (2012).
\bibitem{Poilblanc2013}D. Poilblanc and N. Schuch, Phys. Rev. B {\bf 87}, 140407(R) (2013).
\bibitem{YCHe2013}Y. C. He, D. N. Sheng, and Y. Chen,
Phys. Rev. B \textbf{89}, 075110 (2014).



\bibitem{Bernevig2008}B. A. Bernevig and F. D. M. Haldane, Phys. Rev. Lett. {\bf 100}, 246802 (2008).

\bibitem{Bergholtz2005}E. J. Bergholtz and A. Karlhede, Phys. Rev. Lett. {\bf 94}, 026802 (2005).

\bibitem{Bernevig2012}B. A. Bernevig and N. Regnault, Phys. Rev. B {\bf 85}, 075128 (2012).

\bibitem{Ardonne2005}E. Ardonne, R. Kedem and M. Stone, J. Phys. A:Math. Gen. {\bf 38}, 617 (2005).

\bibitem{Papic}Z. Papic, B. A. Bernevig, and N. Regnault, Phys. Rev. Lett. {\bf 106}, 056801 (2011).

\bibitem{ZLiu2012}Z. Liu, E. J. Bergholtz, H. Fan and A. M. Lauchli, Phys. Rev. B {\bf 85}, 045119 (2012).



\bibitem{Zaletel_flux}M. P. Zaletel, R. S. K. Mong, and F. Pollmann, J. Stat. Mech. 2014, P10007 (2014).

\bibitem{Grushin2014} A. G. Grushin, J. Motruk, M. P. Zaletel, and F. Pollmann,
arXiv:1407.6985.

\bibitem{WZhu_CSL}W. Zhu, S. S. Gong, and D. N. Sheng, arXiv:1410.4883.





\bibitem{ZHWang}E. Rowell, R. Stong, Z. H. Wang, Comm. Math. Phys. {\bf 292}, 343 (2009).
\bibitem{Fendley}P. Fendley, M. P. A. Fisher and C. Nayak, J.Stat.Phys. {\bf 126}, 1111(2007).

\bibitem{Bonderson}P. Bonderson, K. Shtengel and J. K. Slingerland, Phys. Rev. Lett. {\bf 97}, 016401 (2006).

\bibitem{Ardonne}E. Ardonne, E. J. Bergholtz, J. Kailasvuori and E. Wikberg, J. Stat. Mech. 04, 016 (2008).

\bibitem{Bais}F. A. Bais and J. C. Romers, New J. Phys. {\bf 14}, 035024 (2012).


\bibitem{Sandvik}Anders W. Sandvik and G. Vidal, Phys. Rev. Lett. {\bf 99}, 220602 (2007).

\bibitem{Jotzu2014}G. Jotzu, M. Messer,	R. Desbuquois, M. Lebrat, T. Uehlinger,	D. Greif and T. Esslinger, Nature {\bf 515}, 237-240 (2014).

\bibitem{Yao2007}H. Yao, and S. A. Kivelson, Phys. Rev. Lett. {\bf 99}, 247203 (2007).

\bibitem{Ronny2009}M. Greiter and R. Thomale, Phys. Rev. Lett. {\bf 102}, 207203 (2009).

\bibitem{preparation} W. Zhu, S. S. Gong and D. N. Sheng, in preparation. 



\end{thebibliography}
\end{document}